\documentclass{elsart}
\usepackage{graphicx}
\usepackage{bm}
\usepackage{psfrag}
\usepackage{amsmath}
\usepackage{amsfonts}
\usepackage{amssymb}
\usepackage[latin1]{inputenc}

\newcommand{\ket}[1]{|#1\rangle}

\newcommand{\scalar}[2]{\langle#1|#2\rangle}

\newcommand{\op}[1]{|#1\rangle\langle#1|}
\newcommand{\opp}[2]{|#1\rangle\langle#2|}

\newcommand{\intk}{\int_{-\pi}^\pi \frac{dk}{2\pi}}
\newcommand{\intkk}{\intk\int_{-\pi}^\pi \frac{dk'}{2\pi}}

\begin{document}

\begin{frontmatter}
\title{Decoherent quantum walks driven by a generic coin operation}

\author[IFFI]{G. Abal\thanksref{CA}}
\author[IFFI,UFRJ]{R. Donangelo}
\author[IFFI]{F. Severo}
\author[IFFI]{R. Siri}
\address[IFFI]{Instituto de Física, Universidad de la República \\ 
C.C. 30, 11.300, Montevideo, Uruguay}
\address[UFRJ]{Instituto de Física, Universidade Federal do Rio de Janeiro,\\ C.P. 68528, Rio de Janeiro, RJ 21941-972, Brazil}
\thanks[CA]{corresponding author: abal@fing.edu.uy}
\date{\today}

\begin{abstract}
We consider the effect of different unitary noise mechanisms on the evolution of a quantum walk (QW) on a linear chain with a generic coin operation: (i) bit-flip channel noise, restricted to the coin subspace of the QW, and (ii) topological noise caused by randomly broken links in the linear chain. Similarities and differences in the respective decoherent dynamics of the walker as a function of the probability per unit time of a decoherent event taking place are discussed. 
\end{abstract}


\end{frontmatter}

\section{Introduction}

Quantum walks (QW) \cite{Kempe03} are a generalization of random walks to the quantum mechanical regime.  Due to quantum interference effects, the QW spreads faster than its classical counterpart \cite{qw-markov,Tregenna}. This has motivated several optimal quantum search algorithms based on discrete-time \cite{Shenvi,Ambainis05} and continuous-time versions \cite{Childs04} of the QW. Other recent algorithmic applications of quantum walks include the best algorithm for the problem of element distinctness ({\it i.e.} determining if all elements in a set are distinct or not) \cite{Ambainis03} and an algorithm for fast evaluation of NAND trees \cite{Fahri07}. Furthermore, a continuous time quantum walk has been shown to traverse specific binary tree-shaped networks exponentially faster than a classical random walk \cite{Childs02,Childs03}. Even though the significance and generality of some of these results is still under debate \cite{Douglas07}, it is clear that QW's are a highly useful concept for quantum computation. 

Among the many suggestions made for the physical realization of the QW, some are now implemented at the proof-of-principle stage \cite{Ryan,Do-exp}. As in any implementation of a quantum device, the problem of decoherence due to coupling to the environment or to imperfect gate operations, is a mayor obstacle to be dealt with before useful computations can be accomplished using QWs. Quantum error correction protocols require a considerable overhead in quantum resources, so it is important to characterize the effects of decoherence on the dynamics and perhaps even use controlled decoherence to achieve specific purposes \cite{KT03,Tregenna03}. In recent years, several studies on decoherent quantum walks in one  \cite{KT02,Brun03,Biham,Paz03,deco} or more dimensions \cite{Portugal,A+R,Kosik} have appeared.  Ref.~\cite{kendon-review} provides a recent review on the subject. 

The effects of decoherence on the quantum dynamics are frequently studied by applying some quantum operation on the system with probability $p$ per unit time. When the quantum operation is restricted to the coin subspace, considerable progress can be made using analytical techniques. In other cases, path counting-methods have proved to be useful tools, but frequently one must resort to numerical simulation to explore the resulting dynamics in detail. It is a common feature of decoherent quantum walks that for times long compared to a characteristic time $1/p$, the variance increases linearly, but at a higher rate than that implied by classical diffusion. This spreading rate is a useful signature of the effect of decoherence on the quantum walk.

Usually some fixed unitary coin operation is used to drive the quantum walk. However, the impact of decoherence may be different for different coin operations. In this work, a decoherent quantum-walk on the line driven by a generalized coin operation is analyzed. Two different noise mechanisms are considered: (i) a coin-flip operation  applied with probability $p$ per unit time (\textit{{\it i.e.}} bit-flip channel noise) and (ii) at a given time a link on the line is open with probability $p$. This model (broken-link model) was introduced in \cite{deco} as a way to mimic the effects of thermal noise in some experimental situations and it was later generalized to higher dimensions \cite{Portugal,amanda-tesis}. This work is organized as follows: in section~\ref{sec:qwalk} the formalism of the coherent quantum walk is introduced and, in this context, the method of quantum operations is reviewed. In section ~\ref{sec:decoQW} the two specific noise mechanisms mentioned above are considered and their effect on the spreading rate of the quantum walk is discussed. Finally, in section~\ref{sec:conclusions} we present our conclusions.

\section{Generalized discrete-time quantum walk on the line}
\label{sec:qwalk}

The discrete-time quantum walk on the line is a quantum analog of the classical random walk where the random choice is replaced by a unitary operation in the abstract ``coin'' subspace, ${\mathcal H}_C$. This is a single-qubit space, spanned by two orthonormal vectors, usually denoted $\{\ket{R}, \ket{L}\}$. In the one dimensional case, the motion takes place in a position subspace, ${\mathcal H}_P $, spanned by an orthonormal set of position eigenstates $\{\ket{x}\}$ with $x$ an integer associated to discrete positions on a line.  The Hilbert space is then  ${\mathcal H}={\mathcal H}_P\otimes{\mathcal H}_C$ and a generic state for the walker is of the form
\begin{equation}
\ket{\Psi}=\sum_{x=-\infty}^\infty \ket{x}\otimes\left( a_x\ket{R} + b_x\ket{L}\right), \label{sp-wv}
\end{equation} 
where the amplitudes $a_x=\scalar{x,R\,}{\Psi}$ and $b_x=\scalar{x,L\,}{\Psi}$ satisfy the  normalization condition $\protect{\sum_x |a_x|^2+|b_x|^2=1}$. 

A step of the walk is described by the unitary operation
\begin{equation}
U=S_0\cdot\left(I_P\otimes U_C\right) \label{evol1}
\end{equation}
where  $I_P$ indicates the identity in ${\mathcal H}_P$ and $S_0$ is defined by
\begin{equation}
S_0\equiv\sum_x \left(\opp{x+1}{x}\otimes\op{R} + \opp{x-1}{x}\otimes\op{L}\right). \label{shift-x}
\end{equation} 
This unitary operator conditionally shifts the position by one step. The coin operation,  $U_C$, can be any suitable unitary operation in ${\mathcal H}_C$. 

\subsection{Generalized coin operation}

A  parametrization which represents a general unitary operation in ${\mathcal H}_C$ requires three real parameters. For our purposes, a single-parameter coin operation that allows us to explore the response of the system to different environments will suffice. We consider coin operations
\begin{equation}
U_C=
\left(
\begin{array}{cc}
\cos\theta& \sin\theta\\ 
\sin\theta& -\cos\theta
\end{array}
\right)\label{UC}
\end{equation}
parametrized in terms of the real angular parameter $\theta\in[-\pi/2,\pi/2]$. For $\theta=\pi/4$, the QW reduces to the standard Hadamard walk on a line.  

Open quantum systems are best described in terms of density operators. In the absence of noise, an initial state $\rho_0$ evolves, after $t$ iterations, to $\protect{\rho_t=U^t\rho_0 (U^\dagger)^t}$. The probability distribution for finding the walker at site $x$ at time $t$ is $\protect{P(x,t)=tr(\rho_t\op{x})}$, where $tr(\cdot)$ represents a trace operation. It is a well established fact \cite{qw-markov,Travaglione} that the variance of this distribution increases quadratically with time, while in the classical case the increase is only linear.  The quadratic increase is directly related to quantum interference effects and is eventually lost in the presence of decoherence \cite{deco}. 

Since the QW has a constant step size, its dynamics is best described in the Fourier-transformed space \cite{Nayak}.  This subspace, $\tilde{\mathcal H}_k$, is spanned by the Fourier transformed kets $\ket{k}\equiv\sum_x e^{ikx}\ket{x}$, where the real wavenumber $k$ is restricted to $[-\pi,\pi]$.  In this representation the density operator is 
\begin{equation}\label{rho-op}
\rho=\intkk \opp{k}{k'}\otimes\hat\chi_{kk'}
\end{equation}
where $\hat\chi_{kk'}\equiv\opp{\chi_k}{\chi_{k'}}$. In this work we consider pure, localized initial states, and, without loss of generality, start the walker at the origin, $x=0$. The initial state is, therefore, of the form   $\rho_0=\op{\Psi_0}$  with $\protect{\ket{\Psi_0}=\intk \ket{k}\otimes\ket{\chi_0}}$, where $\ket{\chi_0}$ is an  arbitrary initial coin state, with components $\protect{(\tilde a_k,\tilde b_k)^T=(a_0,b_0)^T}$ satisfying $|a_0|^2+|b_0|^2=1$. We use $(a,b)^T$ to indicate a two-component column vector and the amplitudes in Fourier space are $(\tilde a_k,\tilde b_k)^T\equiv\sum_x e^{-ikx}(a_x,b_x)^T$. 

The shift operator is diagonal in $\tilde{\mathcal H}_k$ since $\protect{S_0\ket{k,R}=e^{-ik}\ket{k,R}}$ and\\ $\protect{S_0\ket{k,L}=e^{ik}\ket{k,L}}$. Therefore the evolution operator, eq.~(\ref{evol1}),  is diagonal in $k$ and acts non-trivially in the coin subspace, $U(\ket{k}\otimes\ket{\chi_k})\equiv\ket{k}\otimes U_k\ket{\chi_k}$, with  
\begin{equation}
U_k=\left(
\begin{array}{cc}
e^{-ik}\cos\theta & e^{-ik}\sin\theta \\ 
e^{ik}\sin\theta & -e^{ik}\cos\theta
\end{array} \right).
\label{k-evol}
\end{equation} 

\subsection{Quantum operations}

Long-time effects of different kinds of noise on the position distribution of the walker can be obtained using quantum operations, as described in detail in Ref.~\cite{Brun03}. Here, we review the essential aspects of this method and introduce the appropriate notation. A trace-preserving quantum operation \cite{Nielsen} is described by a set of  Kraus operators $\{A_n\}$, $n=1,2,\ldots N$ assumed to  satisfy
\begin{equation}\label{Kraus-op-cond}
 \sum_{n=1}^N A_n^\dagger A_n=I.
\end{equation}  
Considerable progress in an analytical description of the dynamics is possible when the operators $A_n$ are restricted to the coin subspace ${\mathcal H}_C$. In this case, $\protect{\hat\chi_{kk'}\rightarrow \sum_n A_n\hat\chi_{kk'} A_n^\dagger}$ and, after a single step in the quantum walk, the coin state becomes $\protect{\hat\chi_{kk'}\rightarrow\sum_n U_kA_n \hat\chi_{kk'} A_n^\dagger U_{k'}^\dagger}$. After $t$ such steps are taken, the density operator is of the form (\ref{rho-op}), with
$$
\hat\chi_{kk'}(t)=\sum_{n_1,n_2\ldots n_t} U_kA_{n_t} \ldots U_kA_{n_1}\chi_0 A_{n_1}^\dagger U_{k'}^\dagger\ldots A_{n_t}^\dagger U_{k'}^\dagger.
$$
This transformation is best described in terms of a superoperator ${\mathcal L}_{kk'}$ defined by 
\begin{equation}\label{superop-def}
{\mathcal L}_{kk'}\left[\hat\chi_{kk'}\right]\equiv \sum_{n=1}^N U_kA_n \hat\chi_{kk'} A_n^\dagger U_{k'}^\dagger,
\end{equation}
so that $\hat\chi_{kk'}(t)={\mathcal L}_{kk'}^t\left[\hat\chi_0\right]$. Notice that the diagonal instance of this superoperator, ${\mathcal L}_k\equiv{\mathcal L}_{kk}$, is trace preserving. 

We are interested in the effects of decoherence on the long-time position probability distribution, $P(x,t)$, of the QW. In what follows, unless otherwise stated, the trace operation acts only in the coin degree of freedom. The moments of the position distribution can be obtained  from the above expressions \cite{Brun03},
\begin{eqnarray*}
 \left<x^m\right>&\equiv&\sum_x x^m P(x,t) \\
&=&\sum_x x^m\intkk e^{ix(k-k')} tr\left[\hat\chi_{kk'}(t) \right]\\
&=&(-i)^m\intk \int_{-\pi}^\pi dk'\; \delta^{(m)}(k-k')\, tr\left[\hat\chi_{kk'}(t) \right],
\end{eqnarray*} 
where the sum over positions has been evaluated in terms of derivatives $\delta^{(m)}$ of the delta function. After integration by parts the first two moments, which determine the variance $\sigma^2$, are given by
\begin{eqnarray}
\left<x\right>&=&\intk\sum_{j=1}^t tr\left\{{\mathbf Z}{\mathcal L}_k^j\left[\hat\chi_0 \right]\right\}\qquad\qquad\mbox{and}\label{m1}\\
\left<x^2\right>&=&\intk\sum_{j=1}^t\left[\sum_{j'=1}^j tr\left\{{\mathbf Z}{\mathcal L}_k^{j-j'}\left({\mathbf Z}{\mathcal L}_k^{j'}\left[\hat\chi_0 \right]\right)\right\}\right.\nonumber\\
&&\qquad\qquad\qquad + \left.\sum_{j'=1}^{j-1} tr\left\{{\mathbf Z}{\mathcal L}_k^{j-j'}\left(\left({\mathcal L}_k^{j'}\left[\hat\chi_0\right]\right){\mathbf Z}\right)\right\}\right], \label{m2}
\end{eqnarray} 
where the fact that ${\mathcal L}_k$ is trace-preserving has been used. 

In order to find a clean expression for ${\mathcal L}_k^j\left[\hat\chi\right]$, the reduced density operator $\hat\chi$ is parametrized as a linear combination of Pauli matrices and the identity, $\hat\chi\equiv\sum_{j=0}^3 r_j\sigma_j,$ where $\sigma_0= \mathbf{I}$ is the $2\times 2$ identity and $\sigma_1={\mathbf X},\sigma_2={\mathbf Y}$ and $\sigma_3={\mathbf Z}$ are the usual Pauli matrices. Since $tr(\hat\chi)=2r_0=1$, $r_0=1/2$ is fixed and  ${\mathcal L}_k$ does not affect this component, the state of the walker can be parametrized by the column vector $\protect{\vec R\equiv (r_1,r_2,r_3)^T}$. Then, the action of the superoperator ${\mathcal L}_k$ is described by a $3\times 3$ matrix $M_k$, 
\begin{equation}\label{def-Mk}
\vec R'=M_k\vec R.
\end{equation}
For a specific set of Kraus operators $\{A_n\}$,  the matrix representation $M_k$ can be obtained from eq.~(\ref{superop-def}). We perform an explicit calculation of this kind in the following section.

\subsubsection*{Moments of the position distribution}

In the absence of decoherent events, the (long-time) first and second moments are proportional to $t$ and $t^2$, respectively. In the presence of decoherent noise, the moments of the position distribution may be expressed in terms of  $M_k$ as follows. 

The first moment, eq.~(\ref{m1}), requires the evaluation of $tr\left\{{\mathbf Z}{\mathcal L}_k^j[\hat\chi]\right\}$ in the Pauli representation. Since $\protect{tr\left\{{\mathbf Z} {\mathcal L}_k^j[\hat\chi]\right\}=2r_3'}$, the first moment  reduces to 
\begin{equation}
 \langle x\rangle=(0, 0, 2)\intk\sum_{j=1}^t M_k^j\;(r_1, r_2, r_3)^T.
\end{equation}
Provided the eigenvalues $\lambda_i$ of $M_k$ satisfy $|\lambda_i|<1$, at long times the sum may be approximated by a geometric series which sum is given by the constant,
\begin{equation}
 \langle x\rangle\cong (0, 0, 2)\intk \,G_k\;(r_1, r_2, r_3)^T.
\end{equation}
We have defined the operator
\begin{equation}
G_k\equiv(I-M_k)^{-1}M_k 
\end{equation}
assuming that $I-M_k$, is invertible ($I$ is the $3\times 3$ identity matrix). Note that, due to the scalar product with the row vector $(0, 0, 2)$, only the third row of $G_k$ is relevant. We emphasize that this expression for the first moment of the decoherent evolution is valid only in the limit of very large times. 

A similar expression for the second moment can also be obtained, although the details are more involved. The diagonal term $(j'=j)$ in eq.~(\ref{m2}) is readily evaluated, and represents the ``classical-like'' linear contribution to the variance. The remaining terms represent quantum effects and and  may be regrouped in the form 
\begin{equation}
\langle x^2\rangle=t+\intk\sum_{j=1}^t\sum_{j'=1}^{j-1} tr\left\{{\mathbf Z}{\mathcal L}_k^{j-j'}\left[{\mathbf Z}\left({\mathcal L}_k^{j'}\left[\hat\chi_0 \right]\right)+\left({\mathcal L}_k^{j'}\left[\hat\chi_0 \right]\right){\mathbf Z}\right]\right\}.
\end{equation}
This expression may be written in the $4\times 4$ Pauli representation as 
\begin{equation}\label{m2a}
 \langle x^2\rangle=t+v\cdot\intk\left[\sum_{j=1}^t\sum_{j'=1}^{j-1}
{\mathcal L}_k^{j-j'}(Z_L+Z_R){\mathcal L}_k^{j'}\right]\hat\chi_0,
\end{equation} 
where $\protect{v=(0,0,0,2)}$ and the matrices $Z_L$ and $Z_R$ represent the action of the Pauli operator ${\mathbf Z}$ on the left or right, respectively, {\it i.e.} $\protect{Z_L O\equiv{\mathbf Z}{\mathcal O}}$ and $\protect{OZ_R\equiv{\mathcal O}{\mathbf Z}}$. The explicit form of these matrices is 
\begin{equation}
 Z_L\equiv\left(
\begin{array}{cccc}
0&0&0&1\\
0&0&-i&0\\
0&i&0&0\\
1&0&0&0
\end{array}
\right)\quad
 Z_R\equiv\left(
\begin{array}{cccc}
0&0&0&1\\
0&0&i&0\\
0&-i&0&0\\
1&0&0&0 
\end{array}
\right)
\end{equation}
so $Z_R+Z_L$ is sparse and real. The second moment should not depend on the initial conditions. As shown in Ref.~\cite{Brun03}, this can be made apparent by separating the first component of the initial coin state 
\begin{equation}
 \chi_0=(1/2, 0, 0, 0)^T+(0, r_1, r_2, r_3)^T\label{sep-ci}
\end{equation}
and noting that 
\begin{equation*}
v\cdot {\mathcal L}_k^{j-j'}(Z_L+Z_R){\mathcal L}_k^{j'}(0, r_1, r_2, r_3)^T=v\cdot (1/2,0,0,0)^T=0.
\end{equation*}
After taking into account the action of $Z_L+Z_R$ on the first term of eq.~(\ref{sep-ci}), the expression for the second moment may be further simplified to 
\begin{eqnarray*}
\langle x^2\rangle&=&t+v\cdot \intk\sum_{j=1}^t\sum_{j'=1}^{j-1}
{\mathcal L}_k^{j-j'}(0,0,0,1)^T\nonumber\\
&=&t+(0, 0, 2)\cdot \intk\sum_{j=1}^t\sum_{j'=1}^{j-1}M_k^{j-j'}(0,0,1)^T
\end{eqnarray*}
where the first trivial component has now been omitted. The double series may be summed as before, provided the eigenvalues of $M_k$ satisfy the restriction $|\lambda_i|<1$. Then, in the limit of very large times, the second moment is approximated by 
\begin{equation}\label{m2c}
 \langle x^2\rangle\cong t+(0, 0, 2)\cdot\intk\left[t\,I-(I-M_k)^{-1}\right]G_k\,(0,0,1)^T.
\end{equation}
Thus, the variance of the QW is determined by the element in the third row and column of $G_k$. 

\subsubsection*{Spreading rate}

As we have seen, in the presence of decoherence the first moment is a constant and the second one increases linearly with time. When the rate of decoherent events is small this increase can be considerably faster than the corresponding classical diffusion rate \cite{deco}. We define the spreading rate of a quantum walker as 
\begin{equation}
D_q\equiv\lim_{t\gg 1}\frac{\partial \sigma^2}{\partial t}.\label{sc-def}
\end{equation}  
If the variance $\sigma ^2$ is associated to an ensemble of walkers, the corresponding classical quantity is the diffusion coefficient of a random walk, $D=1$. Since in this work $t$ is a discrete variable, the use of discrete derivatives in the above definition is implied. As mentioned before, the first moment is asymptotically constant and does not contribute to $D_q$. A simple explicit expression for the spreading coefficient can be obtained from eq.~(\ref{m2c}), 
\begin{equation}
 D_q=1+ 2\bar G_{3,3}.\label{sc1}
\end{equation}
The upper bar indicates that an average in $k-$space has been performed, {\it i.e.} $\bar G\equiv\intk G_k$. Thus, the term $\bar G_{3,3}$ is responsible for the faster spreading rate of a decoherent quantum walk relative to a classical walk.  

For any particular kind of noise, eq.~(\ref{superop-def}) allows one to evaluate eq.~(\ref{sc1}) and obtain $D_q$ in terms of the noise rate and coin operation.  In Ref.~\cite{Brun03} this was done for the particular case of a Hadamard walk ({\it i.e.} $\theta=\pi/4$ in eq.~(\ref{k-evol})) and for a set of Kraus operators representing a partial measurement of the coin state with probability $p$ per time-step. In the following, we apply this formalism to other sources of decoherent events and arbitrary coin operations of the form (\ref{UC}).

\section{Decoherent quantum walk}
\label{sec:decoQW}

\subsection{Bit-flip channel}
\label{ssec:spin-flip}

\begin{figure}
 \centering
 \includegraphics[width=10cm]{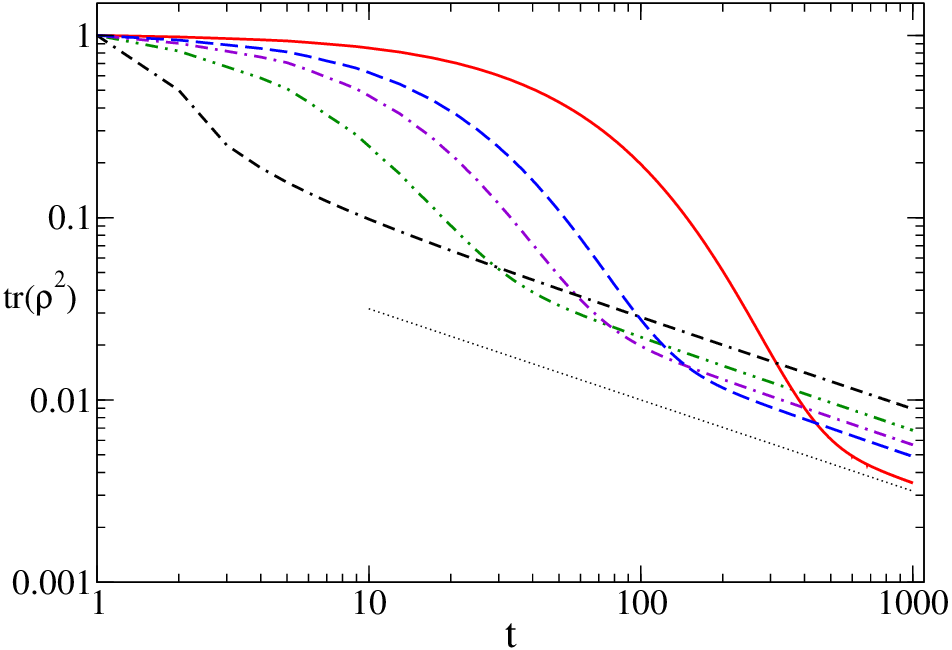}
 \caption{(color online) Purity vs. time for a Hadamard walk $(\theta=\pi/4)$ with bit-flip noise. Several noise rates are shown: $p=0.01$ (full line), $p=0.03$ (dashed), $p=0.05$ (dash-dot), $p=0.10$ (dash-double dot) and $p=0.50$ (dot-double dash). The asymptotic power-law dependence $\sim t^{-1/2}$ is shown as a dotted line to guide the eye.}
 \label{fig:purity}
\end{figure}

Consider the particular set of Kraus operators which flip the coin state with probability $p$ per time step $(p\in [0,1])$,
\begin{equation}\label{bit-flip_Kops}
A_0\equiv\sqrt{p}\;{\mathbf X},\quad A_1\equiv\sqrt{1-p}\; I.
\end{equation}
This particular kind of noise is usually known as \textit{bit-flip channel noise} \cite{Nielsen}. Note that these operators satisfy eq.~(\ref{Kraus-op-cond}) as required from a trace-preserving quantum operation. For the bit-flip channel, eq.~(\ref{superop-def}) reduces to 
\begin{equation}\label{chi-map}
\hat\chi^\prime=p\;U_k{\mathbf X}\,\hat\chi\,{\mathbf X} U_k^\dagger + (1-p)\;U_k\,\hat\chi\, U_k^\dagger
\end{equation}
with $U_k$ given by eq.~(\ref{k-evol}). There are two parameters in this model; the probability per unit time that the coin state is inverted, $p$, and the angle $\theta$ which determines the coin operation. The purity of the state $\rho$, defined as  $\Pi(t)\equiv tr(\rho_t^2)$, can be used as simple indicator of the impact that a given noise rate has on the dynamics. Fig.~\ref{fig:purity} shows the purity as a function of time for several noise rates $p\le 0.50$. For values of $p>0.50$ the purity is identical to that of $1-p$.  For all noise levels and times long compared to $1/p$, the purity decays according to a power law, namely $\Pi\sim t^{-1/2}$. This decay rate was to be expected since, for $t\gg 1/p$, the density operator reduced to the coin subspace, $\rho_C=tr_C(\rho)$, has evolved to a minimum information state, \textit{i.e.} $\rho_C\rightarrow I/2$. In this regime, the position distribution is gaussian-like, peaked at $x=0$, with a characteristic spread $\sigma(p,t)= \sqrt{D_q t}$, as implied by definition~(\ref{sc-def}). A simple calculation shows that this leads to a purity decay $\Pi\sim 1/\sqrt{D_q t}$. 

We now proceed to obtain an explicit expression for the spreading coefficient $D_q(p,\theta)$. After expressing $\chi$ in the Pauli representation introduced in the last section  and using the properties of the Pauli matrices, we obtain from eq.~(\ref{def-Mk}) and eq.~(\ref{chi-map}) the expression for $M_k$,
\begin{equation}
M_k=\left(
\begin{array}{ccc}
 -\cos 2\theta \cos 2k&q\sin 2k &q\sin 2\theta \cos 2k \\
-\cos 2\theta \sin 2k &-q\cos 2k &q\sin 2\theta \sin 2k \\
\sin 2\theta &0&q\cos 2\theta 
\end{array}
\right).
\end{equation}
where $q\equiv 1-2p$. The operator $I-M_k$ may be inverted provided 
\begin{equation}
\det (I-M_k)=(1-q^2)\left[1+\cos 2\theta \cos 2k \right]\neq 0 .
\end{equation}
This condition is satisfied provided $p>0$, $\theta\neq0$ and $\theta\neq\pm\pi/2$. We shall discuss these special coin operations later.

Our main interest is to find the dependence of the spreading coefficient on the parameters $p$ and $\theta$. According to  eq.~(\ref{sc1}), the relevant information is contained in $\bar G_{3,3}$ which equals 
\begin{eqnarray*}
\bar G_{3,3}&=&\frac{q}{1-q^2}\intk \frac{\left(1+q\cos 2\theta \right)\cos 2k 
            +q+\cos 2\theta }{1+\cos 2\theta \cos 2k }\nonumber \\
            &=&\frac{q}{1-q^2}\left[q+\frac{1-|\sin 2\theta |}{\cos 2\theta}\right].
\end{eqnarray*}
Thus, the spreading rate for the case of bit-flip channel noise is,
\begin{equation}\label{Dq-cx}
 D_q(\theta,p)=\frac{1+q^2}{1-q^2}+\frac{2q}{1-q^2}\left[\frac{1-|\sin 2\theta |}{\cos 2\theta}\right].
\end{equation} 
This expression is one of the main results of this work. One of its interesting features is that $D_q=1$ for $p=1/2$, regardless of the coin operation. This can be understood from the fact that $p=1/2$ represents minimum information in eq.~(\ref{chi-map}) and corresponds to the highest decoherence rate. It also explains the fact, apparent in Fig.~\ref{fig:purity}, that lower $p$ values correspond to lower purity at long times due to their faster spreading rates. For $p=0$ and for $p=1$, eq.~(\ref{Dq-cx}) does not apply and in fact diverges. This is to be expected on physical grounds, since in these cases the quantum dynamics is coherent and the variance increases quadratically with time. The case $p=1$, which corresponds to a coherent quantum walk with a modified coin operation $U_C'= U_C X$, has been considered in detail in \cite{Chandra}.

\begin{figure}
 \centering
 \includegraphics[width=8cm]{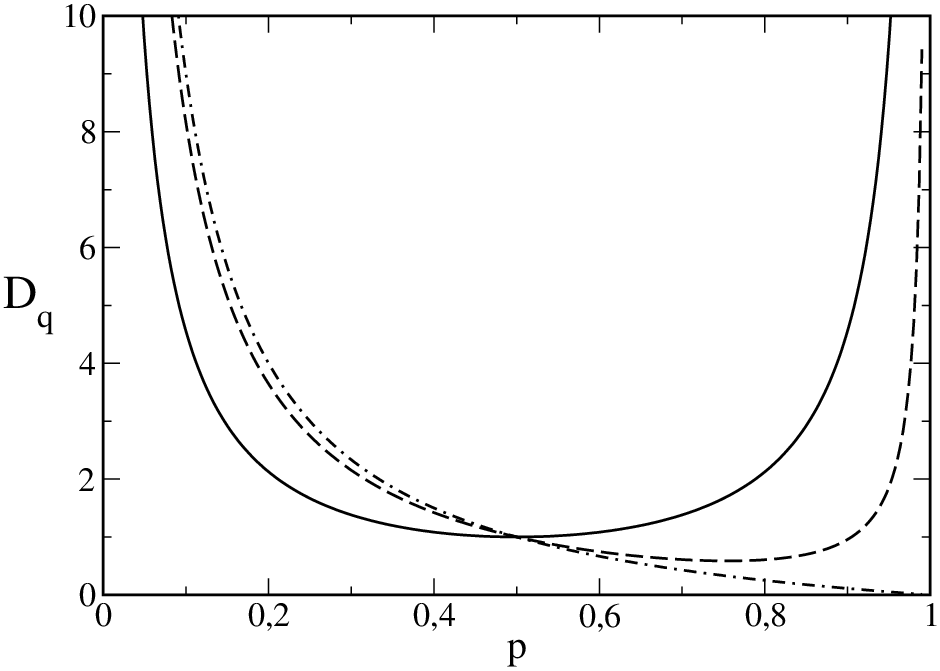}
 \caption{Dependence of the spreading rate $D_q$ on the probability  $p$, from eq.~(\ref{Dq-cx}). Three coin operations are shown: $\theta=\pi/4$  (Hadamard coin, full line), $\theta=\pi/30$ (dashed line) and $\theta=0$, eq.~(\ref{Dq-max}),  (dash-dot line).  }
 \label{fig:Dqp}
\end{figure}

For a Hadamard walk, $\theta=\pi/4$, the second term in the expression for the
spreading rate is null, so (\ref{Dq-cx}) simplifies to, 
\begin{equation}\label{Dq-H}
D_q=\frac{1+q^2}{1-q^2}=\frac{1-2p(1-p)}{2p(1-p)}\qquad\mbox{for~}\theta=\pm\pi/4,
\end{equation}
which for low bit-flip rates, $p\ll 1$, is of the order of $1/(2p)$. The case $\protect{p=1/2}$ corresponds to a minimum spreading rate as shown in Fig.~\ref{fig:Dqp} and higher or lower values of $p$ correspond to lower decoherence and faster dispersion rates. 

For $\theta=0$ the coin operation is diagonal and the quantum walk is equivalent to a classical random walk in which the walker reverses its direction of motion with probability $p$ per time step. This walk has a spreading rate 
\begin{equation}\label{Dq-max}
D_q=\frac{1+q}{1-q}=\frac{1-p}{p}\qquad\mbox{for~}\theta=0.
\end{equation}
For $\theta=\pm \pi/2$, the coin operation reduces to $\bf{X}$ and the quantum walk is equivalent to a classical random walk in which the direction of the walker is preserved with probability $p$. In this case, the spreading rate is
\begin{equation}
D_q=\frac{1-q}{1+q}=\frac{p}{1-p}\qquad\mbox{for~}\theta=\pm \pi/2.
\end{equation}
For small values of $p$, this leads to a very slow spreading rate, a factor of $p^2$ smaller than the maximum rate, eq.~(\ref{Dq-max}).

\begin{figure}
 \centering
 \includegraphics[width=6.7cm]{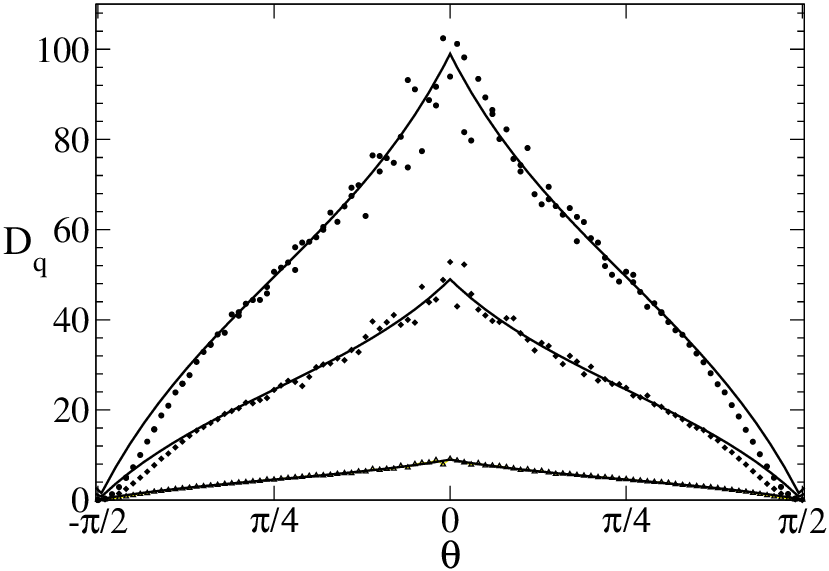}~ \includegraphics[width=6.7cm]{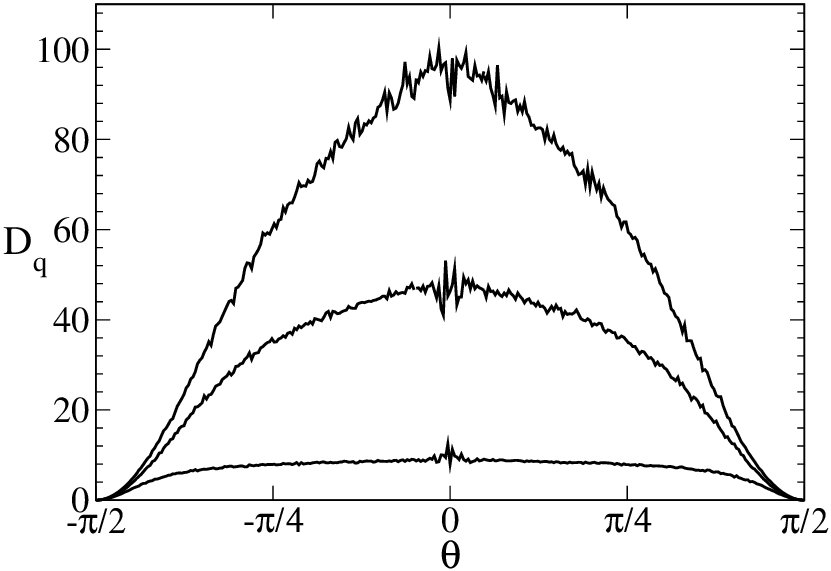}
 \caption{\footnotesize Spreading rate, $D_q$,  for coin operations of the form (\ref{UC}) for noise from different sources. Left panel: Bit-flip channel noise for noise levels $p=0.01, 0.02$ and $0.10$. The full lines correspond to the asymptotic $(t\rightarrow\infty)$ expression for $D_q$, eq.~(\ref{Dq-cx}). The dots are estimates of $D_q$ obtained from eq.~(\ref{sc-def}) through linear regression, after following the time evolution of an ensemble of 1000 walkers for $1000$ time steps. The discrepancy as $\theta$ approaches $\pm\pi/2$ is a consequence of this finite time, as commented in the text. Right panel:  Topological noise (broken links) for the same noise levels as in the right panel. In this case $D_q$ was obtained numerically iterating the map described in Ref.~\cite{deco} for 1000 steps for an ensemble of 1000 walkers. In both cases, the maximum spreading rates are consistent with a $(1-p)/p$ dependence.}\label{fig:Dq-comp}
\end{figure}

We have checked the $\theta$ dependence in eq.~(\ref{Dq-cx}) by performing an independent numerical simulation of the quantum walk on the line in which the coin is inverted with probability $p$. The wavevector at time $t$ results from 
\begin{equation}
 \ket{\Psi(t)}=U_t\ldots U_1\ket{\Psi(0)} \label{U-evol}
\end{equation} 
where each unitary operator $U_j$ is either $U$ with probability $1-p$,\\ or $\protect{U'=(I_P\otimes X)\cdot U}$ with probability $p$. For times long compared with $1/p$, the slope of $\sigma^2$, and thus an estimate for $D_q$, are obtained by linear regression. An ensemble of walkers with the same initial condition was used in order to reduce fluctuations. The resulting dependence of $D_q$ on $\theta$ is shown for several $p$ values in the left panel of Fig.~\ref{fig:Dq-comp} with good agreement with the analytical result. The dispersion rate tends to be  underestimated by the simulation near the singular points $\theta=\pm\pi/2$ due to the fact that the numerical estimate is based on finite-time evolutions. As $\theta$ approaches these singular values, larger times are required to adequately describe the dynamics since the analytical expression corresponds to the limit $t\rightarrow\infty$. We have checked that if larger times are considered, the difference with the theoretical rate is reduced accordingly. 

It is illustrative to compare the spreading rate obtained for bit-flip channel noise with existing results for other noise channels. Noise events can be distinguished according to whether they are restricted or not to the coin subspace of the QW. Among the first kind, Brun et al. \cite{Brun03} obtain analytical results for the dispersion rate of a Hadamard quantum walk (HQW) due to partial coin measurements (or, equivalently, pure dephasing noise). Specifically, they consider the quantum operation defined by $\protect{A_0=\sqrt{2p}\,\op{R}}$, $\protect{A_1=\sqrt{2p}\,\op{L}}$ and $\protect{A_2=\sqrt{1-2p}\, I}$. The spreading rate implied by the results of that work is identical with our eq.~(\ref{Dq-H}). However, it is important to emphasize that the bit-flip channel noise considered here is an example of unitary noise, in the sense of eq.~(\ref{U-evol}), where no measurements are made during the evolution. Shapira \textit{et al.} have previously considered the effect of another kind of unitary noise (also restricted to the coin subspace) on the HQW \cite{Biham}. In that work, in addition to the Hadamard coin operation, a stochastic unitary operator $e^{iA}$ was applied at each step. The hermitic operator  $\protect{A=\alpha_1 X+\alpha_2 Y+\alpha_3 Z}$ was defined at each step by choosing at random the coefficients $\alpha_j(t)$ with variance $\alpha^2=2p$. For small $p$, the numerical results in \cite{Biham} imply that $D_q\rightarrow 1/(2p)$, which is consistent with the weak noise rate limit of eq.~(\ref{Dq-H}). 

A different situation arises when the decoherent events are not restricted to the coin subspace. Kendon and Tregenna have considered both numerically  and analytically \cite{KT03,KT02}  the case of complete measurements (both coin and position) performed with certain probability. They investigate two limiting cases. In the first, the walker takes a number of steps $\ll 1/p$ and those results do not apply to the long time limit considered in this work. The other corresponds to the classical limit, $q\rightarrow 0$. In this later case, their results imply a spreading coefficient  $D_q\approx 1+q^4$, while from eq.~(\ref{Dq-H}) we find $D_q\approx 1+2q^2$, for the case of unitary noise. 

\subsection{Topological noise from broken links}
\label{ssec:broken-links}

A different kind of noise, affecting both the coin and position subspaces, appears when  the links between neighboring sites in the linear chain are broken at random.  The broken-links noise model was introduced in \cite{deco} for the quantum walk on the line and generalized to more dimensions in \cite{Portugal}. It assumes that, at a given time $t$, a site $x$ in the line has one of its neighboring links open with probability $\tilde p$. A broken link remains so for a unit time step, so $\tilde p$ is in fact a probability per unit time. If a link is broken, the corresponding probability flux is not transferred across it and is diverted to the other coin amplitude at the same site. This is done using, in addition to $S_0$ given by eq.~(\ref{shift-x}), the additional shift operators
\begin{eqnarray}\label{Kop}
S_1&=&\sum_x\left(\opp{x+1}{x}\otimes\op{0}+\opp{x}{x}\otimes\opp{0}{1}\right),\nonumber\\
S_2&=&\sum_x\left(\opp{x}{x}\otimes\opp{1}{0}+\opp{x-1}{x}\otimes\op{1}\right),\\
S_3&=&\left(\sum_x\op{x}\right)\otimes\left(\opp{1}{0}+\opp{0}{1}\right)=I\otimes{\mathbf X}.\nonumber
\end{eqnarray} 
During the evolution, each of the operators $S_0,S_1,S_2,S_3$ is applied with probabilities $(1-\tilde p)^2$, $\tilde p(1-\tilde p)$, $\tilde p(1-\tilde p)$ and $\tilde p^2$ respectively, so a set of Kraus operators for this kind of noise is 
\begin{equation}\label{Kraus-bl}
 A_0=(1-\tilde p)\,S,~~ A_1=\sqrt{\tilde p(1-\tilde p)}\,S_1,~~  A_2=\sqrt{\tilde p(1-\tilde p)}\,S_2,~~ A_3=\tilde p\,S_3.
\end{equation} 
This set of operators satisfies eq.~(\ref{Kraus-op-cond}), as required for a trace-preserving quantum operation. The shift operation follows the unitary coin operation $U_C(\theta)$, defined in eq.~(\ref{UC}), so that a step in the evolution can be represented by $\tilde\rho=E\rho E^\dagger$ with 
$$
E\equiv(\sum_{n=0}^3 A_n)\cdot (I_P\otimes U_C).
$$ 
Since the Kraus operators defined by eqs.~(\ref{Kraus-bl}) are not restricted to the coin subspace, the theoretical formalism described in the previous section cannot be applied. 

An alternative approach is to describe the evolution as the result of applying a sequence of uncorrelated unitary operators on the initial state, as in eq.~(\ref{U-evol}). Now each $U_j$ describes the unitary operation that takes place when some specific links in the line are broken at timestep $t=j$. This state--function approach leads to a map for the wavevector amplitudes, as shown in Ref.~\cite{deco}. In that work it was established that for a HQW with randomly broken links the spreading rate depends on $\tilde p$ as 
\begin{equation}
D_q\propto\frac{1-\tilde p}{\tilde p},\label{Dp}
\end{equation}
with a proportionality constant of order 1. A similar dependence has been obtained for higher dimensional quantum walks when the links are broken isotropically \cite{Portugal}. 
Here, we are interested in how $D_q$ depends on the choice of coin operation as determined by the parameter $\theta$. This can be obtained numerically by following the dynamics of an ensemble of walkers to times $t\gg 1/\tilde p$ for several values of $\theta$ and calculating $D_q$ from the $\sigma^2$ vs $t$ data, using eq.~(\ref{sc-def}). Our results, for different noise levels, are shown in the right panel of Fig.~\ref{fig:Dq-comp}. Comparing both panels in this figure, we notice that while both noise models have a similar dependence on the noise rate around $\theta=0$, the details of their behavior in this region are quite different. While the broken-links noise is flat, the bit-flip noise presents a discontinuity in its derivative, specially noticeable at weak noise rates $\tilde p\ll 1$. 

The special cases $\theta=\pm\pi/2$, for which $D_q$ is very small, have been discussed in the previous section. For the case $\theta=0$, where the coin operation reduces to ${\mathbf Z}$, the spreading rate is fastest as was in the case of bit-flip channel noise. In this case the motion of the individual walkers is very sensitive to decoherent events and leads to large fluctuations in the values of $D_q$ obtained through numerical simulation. 
The Hadamard walk corresponds to an intermediate spreading rate. For instance, the HQW spreading rates may be doubled if coin operations with small $\theta$ values are used. 

\section{Conclusions}
\label{sec:conclusions}

The decoherent quantum walk (QW) on the line has been investigated, both analytically and numerically. At  long times compared to the decoherence rate, the QW position distribution spreads out with a variance $\sigma^2$ proportional to the discrete time $t$. The proportionality constant is the  spreading rate, $D_q$, which depends both on the noise rate $p$ and on the particular noise model  considered. The noise models can involve complete or partial measurements or the application of different unitary operations at each time step (unitary noise). In this work, the spreading rate of a generalized QW resulting from two kinds of unitary noise has been investigated: (i)~bit-flip channel noise, restricted to the coin subspace of the QW and (ii)~broken-links noise, which affects both subspaces. 

A family of coin operations generated by a single real parameter $\theta$ has been considered. This  family includes special cases, such as the Pauli operations $X$ and $Z$, and the Hadamard coin, a preferred choice for the one-dimensional QW. The dependence of the spreading rate on this family of coin operations has been investigated for both kinds of noise. In the Hadamard case, $\theta=\pi/4$, our results are compared with existing results for other noise models. The spreading rate is highest for $\theta\approx 0$, which corresponds to essentially diagonal coin operations, close to $Z$. In this case, an initial coin superposition state is preserved throughout the evolution and, in this sense, the walker experiences a simple translation. For coin operations with $\theta\approx\pm\pi/2$, or close to $X$, in the coherent case $(p=0)$ the walker remains ``locked'' in the neighborhood of the starting point and no spreading occurs. For $p>0$, a decoherent event  ``unlocks'' the walker, but the spreading rate is a minimum, independently of the noise model. Due to these general features, the overall dependence of the spreading rate on the coin operation is independent of the noise model. The details are, however, different when noise affects the position subspace and when it does not. 

If the decoherent events are restricted to the coin subspace, considerable progress can be made using quantum operational techniques. For the QW with bit-flip channel noise an analytical expression for the dependence of the spreading rate on the coin operation and on the noise rate has been obtained. This expression, valid for long times, was found consistent with the results of finite-time numerical simulations. The spreading rate increases as $1/p$ for weak noise and, as expected, approaches the classical value $D_q\rightarrow 1$ for high noise rates. Within this noise model, the purity $\Pi=tr(\rho^2)$ of the density operator $\rho$ of the QW was found to decay according to a simple power law, namely $\Pi\sim t^{-1/2}$. For the particular case of a Hadamard walk, the dependence of the spreading rate on the probability $p$ is consistent with the results reported in \cite{Brun03} which uses partial measurements of the coin state as the source of noise. This suggests that the dependence on the frequency of decoherent events may have a generic character or at least be weakly dependent on the details of the quantum operation, when it is restricted to the coin subspace. 

When the noise model affects the position space, the dynamics is different from the previous case. For instance, important differences have been reported in the position distribution using frequent position measurements as a noise model \cite{KT03}. The effects on the spreading coefficient of a different noise model, the broken-link model, which affects both the coin and position subspaces of the QW were investigated and compared to that of the bit-flip noise. The comparison is of interest, because both noise models correspond to unitary noise and involve no measurements, but one of them (bit-flip) is restricted to the coin subspace. A different dependence on the coin operation parameter $\theta$ was found, but the dependence on the noise level parameter $p$ is similar in both cases, particularly at low noise levels.

The previous considerations may be summarized by noting that the noise rate $p$ is of  primary importance in determining the dynamics of the decoherent QW. In all cases considered, the spreading rate is higher than the corresponding classical rate by a factor of the order of $1/p$. In a previous work \cite{deco} we argued that this is an indication of persistent quantum correlations. In other words, even for arbitrarily long times, partial quantum coherence continues to play a role in the dynamics of the evolution. A similar conclusion is obtained for the coin-flip noise introduced in the present work.

{\it G.A. and R.S. acknowledge support from PEDECIBA and PDT (Proy. No.54). R.D. and G.A. acknowledge support from the Milenium Institute for Quantum Information.} 

\bibliography{qw}
\bibliographystyle{h-elsevier}
\end{document}